\documentclass[pre,twocolumn,floatfix,,amsmath,showpacs]{revtex4}
\usepackage[dvips]{graphicx}
\usepackage{amsfonts}

\begin{document}

\title{Transport properties of a modified Lorentz gas}

\author{H. Larralde, F. Leyvraz and C. Mej\'\i a--Monasterio}
\affiliation{Centro de Ciencias F\'\i sicas, Universidad Nacional Aut\'onoma de Mexico \\
Av.~Universidad 1001, Colonia Chamilpa, Cuernavaca, Morelos 62210 Mexico}
\date{\today}

\begin{abstract}
  We present a detailed study of the first simple mechanical system that shows
  fully realistic transport behavior while still being exactly solvable at the
  level of equilibrium statistical mechanics.  The system under consideration
  is a Lorentz gas with fixed freely-rotating circular scatterers interacting
  with point particles via perfectly rough collisions. Upon imposing a
  temperature and/or a chemical potential gradient, a stationary state is
  attained for which local thermal equilibrium holds for low values of the
  imposed gradients. Transport in this system is normal, in the sense that the
  transport coefficients which characterize the flow of heat and matter are
  finite in the thermodynamic limit. Moreover, the two flows are non-trivially
  coupled, satisfying Onsager's reciprocity relations to within numerical
  accuracy as well as the Green-Kubo relations .  We further show numerically
  that an applied electric field causes the same currents as the corresponding
  chemical potential gradient in first order of the applied field. Puzzling
  discrepancies in higher order effects (Joule heating) are also observed.
  Finally, the role of entropy production in this purely Hamiltonian system is
  shortly discussed.
\end{abstract}

\maketitle


\section{Introduction}
\label {sec:intro}

One of the major motors of research in statistical physics has been the quest
to understand the link between macroscopic phenomena and the underlying
microscopic physics of a system. In particular, the origin of the macroscopic
``laws'' of thermodynamic transport is still one of the major challenges to
theoretical physics. These phenomenological laws are known to describe
accurately the processes of diffusion, heat conduction, viscosity among a host
of other phenomena, and are fundamental to the quantitative description of
macroscopic systems in general. However, the attempts to link these
macroscopic laws to the underlying microscopic dynamics have not been
conclusive thus far.  From a mathematically rigorous point of view, very few
results have been obtained \cite{bonetto01}. Indeed, to our knowledge, the
validity of Fourier's law has been proven analytically only for a very
specific model in the limit of infinite dilution with finite mean free path
\cite{lebowitz78,lebowitz82}. There have also been attempts to link transport
phenomena to the chaotic properties of the underlying classical dynamics
\cite{bunimovich80}; and a connection between the rate of entropy production
and the rate of contraction of phase space volume in thermostated (non
Hamiltonian) systems has been pointed out \cite{chernov93, holian87, moran87,
  chernov93-2}.

Given this state of affairs, a common strategy is to propose and study systems
which reproduce, in numerical simulations, the phenomena under consideration,
and to attempt to determine how the physical ingredients of these models give
rise to the macroscopic behaviour. However, the systems considered thus far
have been either too complicated to shed much light upon the problem, or have
actually failed to reproduce the macroscopic phenomenology. Attempts have been
made, on one side, through the simulation of realistic many-body systems
satisfying a thermostated dynamics (see for example \cite{evans86}). These
simulations have indeed been able to reproduce non-trivial transport
phenomena. However, such studies do not provide a detailed understanding of
the microscopic processes involved due to the excessive complexity of the
system under study. The other numerical approach involves the study of
transport in ``simple systems''. Examples of these include: chains of
anharmonic oscillators \cite{lepri97} and the so called ding-a-ling and
ding-dong models \cite{casati84,prosen92}, among others. Of these, energy
transport in the chains was shown to be ``anomalous''. This is an euphemism
indicating that energy transport cannot be described as a diffusive process
and currents do not scale with the gradients in the expected way. In
contrast, ``normal'' transport indicates that even in the thermodynamic limit,
transport can accurately be described as a diffusive process and that the flux
is proportional to the gradient as stated, say, in Fourier's law. The
ding-a-ling and ding-dong models, indeed do yield normal transport under
certain conditions; however, they include geometric constraints that make even
their equilibrium properties an extremely complicated affair. As we shall see,
the knowledge of such equilibrium properties is often useful, which is why we
consider it important to have a model where these are explicitly known.

One particularly thorny problem in attempting to reproduce the
phenomenological laws of thermodynamic transport, is that these apply to
systems which are characterized by local thermodynamic equilibrium (LTE)
\cite{rondoni02,cohen02}. That is, systems that are not in equilibrium, but
for which the intensive thermodynamic variables are well defined at each point
of the system, and the relations amongst these variables are the same as in
equilibrium thermodynamics. Thus, results have been obtained concerning
``diffusive'' (normal) energy transport in the Lorentz gas \cite{alonso99}.
Yet this system is not described by LTE \cite{dhar99}. It is therefore not
clear what precise meaning can be attached to the local temperature appearing
in Fourier's law in this situation.

Under these circumstances, the minimal ingredients required in the microscopic
physics of a simple system to attain normal thermodynamic transport in low
dimensions are presently still under discussion
\cite{bonetto01,hu98,prosen00}. However, we believe that ideal candidate
models should possess the following features:
\begin{enumerate}
\item The microscopic physics should be defined in terms of reversible
 Hamiltonian dynamics, since this is the nature of known fundamental
 processes. In particular, reversible systems showing an average rate of
 phase space contraction, while of considerable interest for simulating
 transport, do not provide a fundamental microscopic model.
 
\item The equilibrium properties of the model should be well understood.
 
\item When driven weakly out of equilibrium, they must be consistent with the
 hypotheses of LTE and, of course, they must give rise to realistic
 macroscopic transport.
\end{enumerate}

In this work we present a detailed description and extensive simulations of
the equilibrium and transport properties of a simple reversible Hamiltonian
model system which we believe is an ideal candidate to begin to unravel the
puzzle of thermodynamic transport. The first results concerning some of the
transport features of this model were presented in \cite{mejia01}. In
Section~\ref{sec:model}, we present a detailed description and discussion of
the model. We show that its equilibrium statistical mechanics is simply that
of an ideal gas.

In order to carry out the equilibrium simulations, as well as to study the
transport properties of the model when the system is driven out of
equilibrium, it is necessary to couple the system to thermo-chemical baths. We
describe the model baths we have implemented to do this in
Section~\ref{sec:baths}.

In Section~\ref{sec:lte} we corroborate through simulations that the
equilibrium state of the system in the three canonical ensembles is indeed an
ideal gas. We also present numerical evidence to show that, when subjected to
a weak temperature and/or chemical potential gradient, our systems reaches a
non-equilibrium steady state (NSS) which is characterized by LTE. In this
situation, our model supports both heat and matter flows. We show in
Section~\ref{sec:transport} that these flows are characterized by transport
coefficients which are independent of system size (that is, transport is
normal), and that the corresponding transport coefficients satisfy Onsager's
reciprocity relations. The fact that the system is homogeneous in energy
allows the prediction of the dependence of the transport coefficients on the
temperature. We also establish numerically their dependence on the particle
density and discuss briefly the effects of subjecting the system to a magnetic
field. In Section~\ref{sec:green-kubo} we verify that the Green-Kubo formulas
connecting transport coefficients with time correlation functions apply
for this system to within numerical precision, and we compare the effect of an
applied electric field with that of an applied chemical potential gradient,
which, to linear order, should induce the same flows in the system. In
Section~\ref{sec:entropy} a brief discussion of the applicability of
microscopic interpretations of entropy production as motor of transport in
this system. Finally we present a brief summary and mention how this model can
be modified to study other physical transport problems from a microscopic
approach.


\section{Definition of the model}
\label {sec:model}

In \cite{alonso99}, a Lorentz gas model with elastic collisions was used to
study heat transport in a quasi-one dimensional channel placed between two
thermal reservoirs at different nominal temperatures.  While the results were
consistent with some sort of ``diffusive'' energy transport, identification
with Fourier's law was unfounded: the system does not satisfy the hypothesis
of LTE and, therefore, one cannot define a local temperature, as was shown in
\cite{dhar99}.  There, the authors argue that the system does not attain LTE
due to the existence of an infinite number (in the thermodynamic limit) of
conserved quantities in the dynamics. Indeed, the energy of each particle is
conserved throughout the evolution of the system. This, in turn, implies a
breakdown of ergodicity and the resulting process is closer to ``colour''
diffusion than to heat transport.

The model we study in this work, introduced in \cite{mejia01}, is a
modification to the usual Lorentz gas model in which the scatterers are
allowed to exchange energy with a set of (non-interacting) point particles of
mass $m$ through the scattering events. The geometry we consider is that of a
periodic Lorentz gas in which the hard disc scatterers of radius $R$ are fixed
on a triangular lattice, the details of which are discussed below. The
possibility of energy exchange is achieved as follows: each disc is a free
rotator with a moment of inertia $\Theta$, and scattering proceeds according
to rules characterizing ``perfectly rough'' collisions, that are reversible,
conserve total energy and angular momentum. These collision rules are given by
the following formulas, which relate the normal and tangential components of
the particle's velocity ${\bf v}$ with respect to the disc's surface, and the
disc's angular velocity $\omega$ before (unprimed quantities) and after
(primed quantities) the collision
\begin{eqnarray}
v_n' & = & -v_n\, \nonumber\\
v_t' & = & v_t - \frac{2\eta}{1 + \eta} (v_t - R\omega)\, \\
\label{eq:collision}
R\omega'& = & R\omega + \frac{2}{1 + \eta} (v_t - R\omega) \nonumber \ .
\end{eqnarray}

These rules define a deterministic, time-reversible, canonical transformation
at each collision. The parameter $\eta$, defined as the ratio between the
moment of inertia of the disc and the mass of the particle times the square
radius of the disc:
\begin{equation}
\eta=\frac{\Theta}{mR^2} \ ,
\label{eq:defeta}
\end{equation}
is the only relevant adimensional parameter characterizing the collision. It
determines the energy transfer between discs and particles in a collision. For
finite values of $\eta$, particles in the system may exchange energy among
each other through the discs, even though they do not interact directly. This
simple energy exchange mechanism overcomes the objections raised in
\cite{dhar99} and permits the system to reach thermodynamical equilibrium, as
we will see in Section~\ref{sec:lte}. As $\eta \to 0$, the rotational energy
of the scatterers becomes negligible and the collision becomes perfectly
elastic, recovering the dynamics of the usual Lorentz gas model. For $\eta \to
\infty$, the energy of the scatterers is unaffected by the collisions with the
particles, and the angular velocity of the scatterers remains constant. In
this limit, once again, the system does not equilibrate. Thus, in either
limit, the energy-mediating effect is suppressed and thermodynamical
equilibrium is not reached. In the following, unless the contrary is
explicitly stated, we shall always be dealing with the case $\eta = 1$, since
that is a value of $\eta$ for which energy exchange, and therefore
equilibration, is quite efficient.

\begin{figure}[!t]
\begin{center}
\includegraphics[scale=0.4]{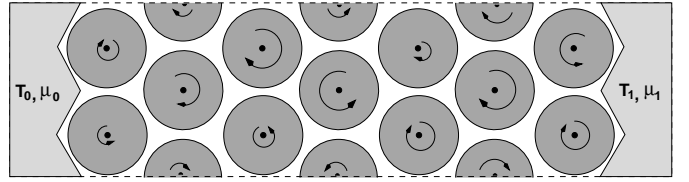}
\caption{\label{fig:channel}Schematic illustration of the scatterer
geometry: the scatterers are disposed on a triangular array with
finite horizon to avoid infinitely long trajectories. For matters of
convenience, in this work the separation between scatterers has been
set to have the critical horizon. Periodic boundary conditions are
used in the vertical direction. To avoid spurious effects arising
from multiple consecutive scatterings of a particle off the same disc,
we have put two discs on each vertical. To study the dependence with
system size, the length L of the sample is varied. The quoted length
is the number of discs.}
\end{center}
\end{figure}

The geometric disposition of the scatterers in the systems we study in this
work is indicated in Fig.~\ref{fig:channel}. The centers of the scatterers are
fixed on a triangular lattice, along a narrow channel of height $2W$, where
$W$ is the distance between the centers of the scatterers. For convenience,
this distance was chosen as $W=4R/\sqrt{3}$ (known in the literature as the
critical horizon), which is the largest separation for which a particle cannot
travel arbitrarily large distances without undergoing a collision. In this
geometry, a set of non-interacting point particles of mass $m$ moves freely
between collisions with the hard discs. In the vertical direction the channel
contains two discs and periodic boundary conditions are imposed.

The fact that we put two discs in the vertical direction merits comment as, in
the simpler ``single cell'' geometry with periodic boundary conditions,
spurious effects may arise from multiple successive scatterings of one
particle with the same disc. Some of these effects have been observed in
\cite{klages00}, where the same set of collision equations has been used to
model a deterministic thermostat. We have verified that if one considers a
system consisting of one cell with periodic boundary conditions, containing a
single scatterer and a single particle, the resulting dynamics gives rise to
regular structures in the phase space of the particle. In this situation, the
system does not appear to be ergodic for arbitrary values of $\eta$. We also
noted that, if instead of periodic boundary conditions one considers specular
reflections at the boundaries, the effect in the particle's trajectory is a
``randomization'' of the tangential component of the particle's velocity in
the next collision with the disc, recovering a seemingly ergodic phase space.
Though we believe that the presence of other particles will destroy the
regular structures that appear in the single particle case, we decided to
avoid the possibility of multiple consecutive collisions of a particle with
the same disc by placing two discs in the vertical direction.

A central aspect of this model is that its equilibrium properties are still
trivial, even though, strictly speaking, it is an interacting many particle
system. Indeed, in spite of the modified collision rules, the energy of the
system is given by
\begin{equation}
E=m\sum\limits_{particles} \frac{{\bf v_i}^2}{2} + \Theta\sum\limits_{rotors} 
\frac{\omega_\alpha^2}{2} \ .
\end{equation}

Thus, all statistical mechanics calculations for this system coincide with
those of a system consisting of a two dimensional ideal gas plus a collection
of non-interacting free rotors. Hence, a first test for the system is to
verify that it equilibrates to a state in which its statistical properties are
indeed those predicted by equilibrium statistical mechanics. In particular,
in microcanonical simulations, the particle velocities should reach Maxwellian
distributions at a uniform ``temperature'' consistent with the equipartition
theorem \cite{homogeneity}. These temperatures should also characterize the
distribution of angular velocities of the rotating scatterers. The same should
be true in canonical and grand canonical simulations, where the temperature and
particle density (or, more formally, the chemical potential divided by the
temperature) are now those established by the values of the baths.
Furthermore, the equations of state characterizing ideal gases should describe
the thermodynamics of our system. If any of these tests fails, it could be
argued that the system fails to equilibrate probably due to a lack of
ergodicity. Unfortunately, passing all the tests does not prove that the
system is ergodic. We have not succeeded in showing rigorously that this
system is ergodic. However, these numerical tests are fairly stringent: we
shall see that for the case of an imposed external magnetic field, where
ergodicity is known to be violated, an effect indeed appears in the energy
distribution function of the particles.

\section{Thermo-chemical Baths}
\label {sec:baths}

In numerical studies of statistical mechanics systems in general, and in those
on the microscopic origin of macroscopic transport in particular, it is
necessary to define models for the thermodynamical reservoirs in addition to
the dynamics of the system. The design of these reservoirs and the way these
interact with the system, has several subtleties that can produce confusing
and, very often, wrong results, as has been pointed out in
\cite{koplik98,hatano99}. In this Section, we explain the design of a
stochastic model that simulates a thermo-chemical bath, which is able to
exchange energy and particles with the system at fixed nominal values for the
temperature $T$ and chemical potential $\mu$.

The heat and matter reservoir is taken to be an infinite ideal gas at
temperature $T$ and density $\rho$ which is placed in direct contact with the
system being studied. This is achieved by assuming that the wall separating
the system from the reservoir is perfectly transparent to particles impinging
on it from either side. Of course, it is not necessary to simulate explicitly
the infinite ideal gas which acts as the thermodynamical reservoir, instead,
we can implement this set up using the following rules: whenever a particle of
the system impinges on the boundary which separates it from the bath, it is
removed. On the other hand, with a frequency $\gamma$, particles are generated
at the boundary, with a velocity distribution
\begin{eqnarray}
P_n(v_n) & = & \ \frac{m}{T}|v_n|
\exp\left(-\frac{m v_n^2}{2T}\right) \ , \nonumber\\
P_t(v_t) & = & \sqrt{\frac{m}{2\pi T}}
\exp\left(-\frac{m v_t^2}{2T}\right) \ ,
\label{eq:dist}
\end{eqnarray}
reflecting the assumption that the bath is an ideal gas (here and in the
following, we take the particle mass and Boltzmann's constant equal to one).
The choice of $T$ in these equations defines the nominal temperature of the
bath. Moreover, this way of implementing the thermo-chemical baths also fixes
a nominal value for their chemical potential. Again, as the bath is an ideal
gas, the rate $\gamma$ is given by:
\begin{equation} \label {eq:gamma}
\gamma = \frac{1}{\sqrt{2\pi}}\rho~T^{1/2} \ .
\end{equation}
Thus, we can express the chemical potential of the bath in terms of the
parameter $\gamma$ as:
\begin{equation} \label {eq:mu}
\mu = T\ln\left(\frac{\lambda_0 \gamma}{T^{3/2}}\right) \ ,
\end{equation}
where $\lambda_0$ is an irrelevant constant.

Equations (\ref{eq:dist}), (\ref{eq:gamma}) and (\ref{eq:mu}) completely
define the algorithm for the stochastic emission process of the
thermo-chemical baths used in the simulations, and allow us to control the
chemical potential and temperature of the walls by varying the rate $\gamma$
and the temperature. These walls were used for grand canonical simulations in
equilibrium and coupled heat and matter transport simulations in the NSS.

To simulate the canonical ensemble in equilibrium and pure heat flow in the
NSS, diathermal impermeable walls are required. These were achieved by
reflecting each particle that impinged on the wall back into the system, with
its velocity updated according to the distribution function (\ref{eq:dist}),
which again defines the nominal temperature of the thermal bath.

Finally, in order to perform simulations in the microcanonical ensemble,
insulating walls are required. We have simulated these either by simply
considering that the particles always perform elastic specular reflection at
the boundary, or by having no walls at all and considering periodic boundary
conditions.

A generalized model for thermo-chemical bath, as well as a general approach
allowing to verify the validity of a given procedure for generating such baths
is given in Appendix \ref{app-A}.


\section{Equilibrium and Local thermal equilibrium}
\label {sec:lte}

In this section we show that the system described in Section~\ref{sec:model}
reaches a well defined equilibrium state for the different equilibrium
ensembles. For each ensemble we place the system in contact with the
appropriate wall, as described in the previous Section. However, as expected,
the equilibrium state that the system reaches does not depend on the
particular choice of the ensemble. Furthermore, when the nominal values for
the thermodynamical quantities fixed by the baths impose a gradient in
temperature and/or in the chemical potential, our model reaches a well defined
steady state.

\begin{figure}[!t]
\begin{center}
\includegraphics[scale=0.4]{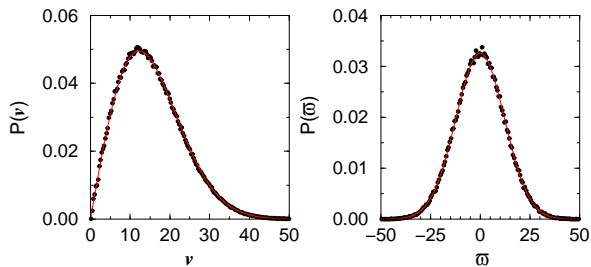}
\caption{\label{fig:equi-dists}Discs angular velocity distribution
$P(\omega)$ and particles velocity distribution $P(v)$ obtained in a
microcanonical simulation with $30$ particles in a channel of length
$L=30$. The solid lines correspond to the Boltzmann distribution at
the expected temperature $T=150$.}
\end{center}
\end{figure}

In order to show that our model reaches a satisfactory equilibrium state we
have measured the velocity distribution of the particles $P(v)$ and the
angular velocity distribution of the discs $P(\omega)$ in a microcanonical
simulation in which the particles undergo specular reflections with the walls
at the left and right extremes of the channel. We have fixed the energy of the
system and distributed it randomly among discs and particles. After some
relaxation time the mean energy per particle is twice the mean energy per
disc, thus satisfying the equipartition theorem. In Fig.~\ref{fig:equi-dists},
the measured distributions $P(v)$ and $P(\omega)$ are shown. The solid line is
the corresponding Boltzmann distributions at the expected temperature ($T=150$
in arbitrary units). The agreement indicates that both discs and particles
have reached a state consistent with the predictions of equilibrium
statistical mechanics. This agreement also allows us to relate the average
energy per particle with the temperature, which would have been unfounded
otherwise.

We have computed the particle density and temperature profiles in the
following way: We divide the channel in ${\mathcal R}$ disjoint slabs of width
$\Delta x = L/{\mathcal R}$. As before, $L$ is the length of the channel in
the $x$-direction. The particle density, $n(x)$, in each slab is computed as
the time average
\begin{equation} \label{eq:concentration}
n(x)dx = \frac{1}{{\mathcal T}} \int_0^{\mathcal T} \sum_{i=1}^N
\delta(x-x_i(t))\,dt\,dx \ ,
\end{equation}
where $x_i(t)$ is the position of the $i$-th particle at time $t$ and
$N$ is the total number of particles in the channel that, in the case
of a grand canonical situation, depends on time. The time average is
performed after the steady state has been reached.

Similarly, the time averaged energy density $\varepsilon(x)$ is
computed as
\begin{equation} \label{eq:energy}
\varepsilon(x)dx = \frac{1}{{\mathcal T}} \int_0^{\mathcal T}
\sum_{i=1}^N E_i(t)\delta(x-x_i(t))\,dt\,dx \ .
\end{equation}

Here $E_i(t)$ is the energy of the $i$-th particle at time $t$. From
(\ref{eq:concentration}) we obtain the particle's density profile
$\rho(x)$ as the mean number of particles found in the slab which
contains the position $x$
\begin{equation} \label{eq:density-profile}
\rho(x) = \int_{\Delta x} n(x)\,dx \ .
\end{equation}

The integral in (\ref{eq:density-profile}) is taken over the domain of
the slab that contains position $x$. Analogously, the average particle
energy $E(x)$ at the slab containing position $x$ is given by
\begin{equation} \label{eq:energy-profile}
E(x) = \int_{\Delta x} \varepsilon(x) {\mathrm d}x \ .
\end{equation}

\begin{figure}[!t]
\begin{center}
\includegraphics[scale=0.4]{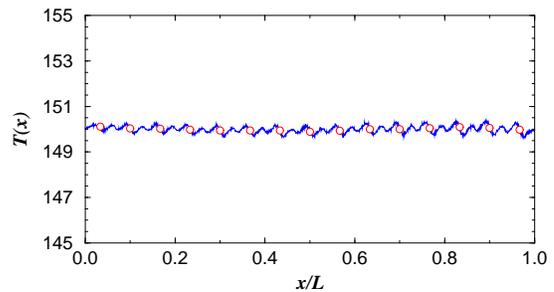}
\caption{\label{fig:equi-T-profile}Particle's temperature profile
$T(x)$ for a canonical simulation. The number of particles was $N=30$
and $L=30$. The nominal temperature in both baths was set to same
value $T=150$. The open circles correspond to the time averaged
temperature of the discs.}
\end{center}
\end{figure}

With (\ref{eq:density-profile}) and (\ref{eq:energy-profile}), we
calculate the particle temperature profile $T(x)$ as
\begin{equation} \label{eq:temperature-profile}
T(x) = \frac{E(x)}{\rho(x)} \ .
\end{equation}
In Fig.~\ref{fig:equi-T-profile}, we show the temperature profile obtained in
a canonical simulation where both baths were set to the same nominal
temperature. The temperature obtained according to
(\ref{eq:temperature-profile}) has additionally been averaged over an ensemble
of $500$ different realizations. The number of particles was set to $N=30$ in
a channel of length $L=30$. We observe that the particles reach an equilibrium
state characterized by a constant temperature along the channel which
coincides with the nominal values of the baths' temperature. Moreover, the
open circles in Fig.~\ref{fig:equi-T-profile} correspond to the time averaged
energy of the discs. The agreement of both profiles indicates equilibration
between particles and discs.

In this equilibrium state, the particle's density profile (not shown), is also
flat. The same behaviour was also obtained in grand canonical simulations, as
was to be expected from the equivalence of the different statistical
ensembles.

\begin{figure}[!t]
\begin{center}
\includegraphics[scale=0.4]{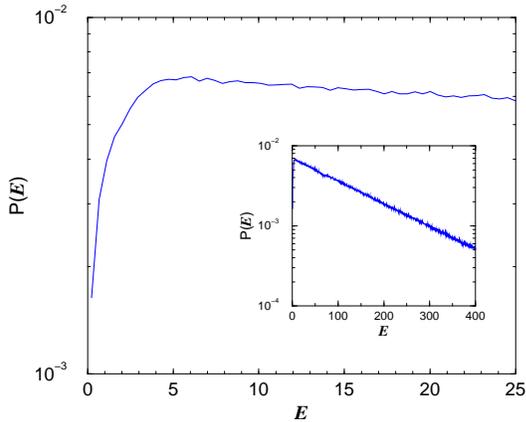}
\caption{\label{fig:magnetic}Particle's energy distribution obtained
from a microcanonical simulation with an applied external magnetic
field. The magnitude of the field $B$ is such that $qB/mc = 10$, where
$q$ and $m$ are the electric charge and mass of the particles
respectively and $c$ is the speed of light. The lack of low energy
particles is evident. In the inset, ${\mathrm P}(E)$ is shown on a
larger energy domain.}
\end{center}
\end{figure}

Let us finally note the  following: we have also performed simulations
involving an external magnetic field.  In this case, it is immediately
clear that  the system  is not ergodic.  Indeed, there  exist isolated
circular orbits  which do not touch  any disc. Since  particles do not
interact directly, any particle originally on one of these orbits will
remain for ever. Similarly, this  set of orbits cannot be reached from
initial conditions for  which each particle touches one  disc at least
once. Of  course, for small fields,  the orbits that do  not touch any
scatterer only  occur when the  particle's kinetic energy is  low, but
since the  particles can exchange  energy with the discs,  the kinetic
energy of  a particle  can come arbitrarily  close to zero,  and there
will be  regions that become  unreachable for this particle.  There is
thus  a true lack  of ergodicity  for this  system for  arbitrary (non
zero)  magnetic fields and  at all  temperatures, although  the effect
becomes  weaker   as  the  field   decreases.   We  have   plotted  in
Fig.~\ref{fig:magnetic} the  distribution of particle  energies in the
microcanonical  ensemble  and  clear  deviations  from  the  Boltzmann
distribution are observed. This comes  as an indication that the tests
we have  applied should  disclose a lack  of ergodicity in  the system
without magnetic field if such were present.

\begin{figure}[!t]
\begin{center}
\includegraphics[scale=0.4]{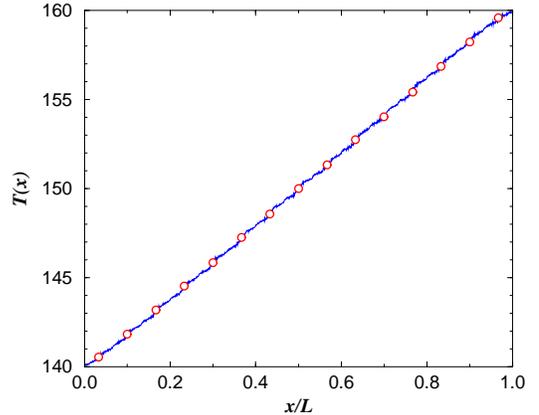}
\caption{\label{fig:T-profile} Temperature profile $T(x)$ with
temperature and chemical potential gradients. The nominal values for
the baths' temperature are $T_0=140$ and $T_1=160$ with an end-to-end
chemical potential difference of $\Delta(\mu/T) = -0.2$. The mean
number of particles was $\approx 25.2$ in a channel of length $L=30$.
The open circles correspond to the mean kinetic energy of the
scatterers.}
\end{center}
\end{figure}

We now turn to the more interesting out of equilibrium situation. In these
simulations we connect the two ends of the system with two different baths,
each of which has given values of the chemical potential and the temperature.
To drive the system out of equilibrium, the nominal values of the temperatures
and chemical potentials of these baths are set to differ by fixed amounts.
Under these conditions, the system is allowed to evolve until a
non-equilibrium steady state (NSS) develops. In Fig.~\ref{fig:T-profile}, we
show results for a typical simulation; the profile corresponds to the average
energy per particle for discs and particles obtained in a channel of length
$L=30$. The temperature difference of the baths was set to $\Delta T=20$
around a central value $T=150$, with a chemical potential difference of
$\Delta(\mu/T)=-0.2$.  The profile of the average energy per particle is
linear and coincides at the boundaries with the nominal baths' temperature. As
in equilibrium, in the steady state discs and particles locally equilibrate to
the same value of the local mean energy per particle, giving a first
indication of the establishment of LTE. As we will show further on, we will be
justified in identifying the mean kinetic energy per particle with the local
temperature.

\begin{figure}[!b]
\begin{center}
\includegraphics[scale=0.4]{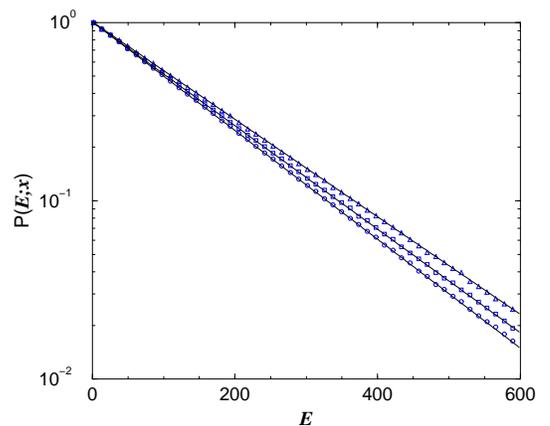}
\caption{\label{fig:lte}Semi-logarithmic plot of the particle's energy
distribution $P_x(E)$ at different positions along a channel obtained
from the simulation described in Fig.~\ref{fig:T-profile}. The
different curves correspond to a fit to a Boltzmann distribution for
each position. From these fits, we obtain the temperatures: $T =
142.69$ at $x/L = 0.1333$ (circles), $T = 149.70$ at $x/L = 0.4666$
(squares) and $T = 158.69$ at $x/L = 0.9333$ (triangles). The curves
have been re-scaled for clarity.}
\end{center}
\end{figure}

\begin{figure}[!t]
\begin{center}
\includegraphics[scale=0.4]{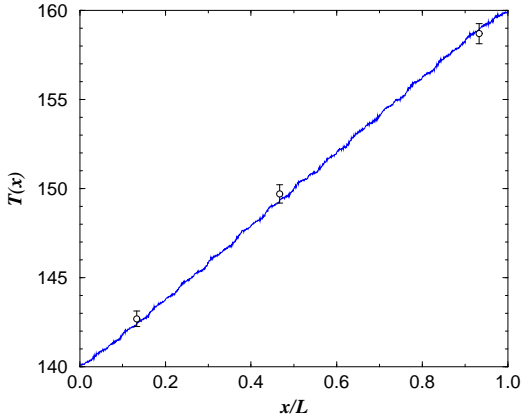}
\caption{\label{fig:lte-2} Temperature profile of
Fig.~\ref{fig:T-profile} compared with the temperatures (open circles)
obtained from the fit to the Boltzmann distribution of $P(E;x)$ shown
in Fig.~\ref{fig:lte}.}
\end{center}
\end{figure}

\begin{figure}[!b]
\begin{center}
\includegraphics[scale=0.4]{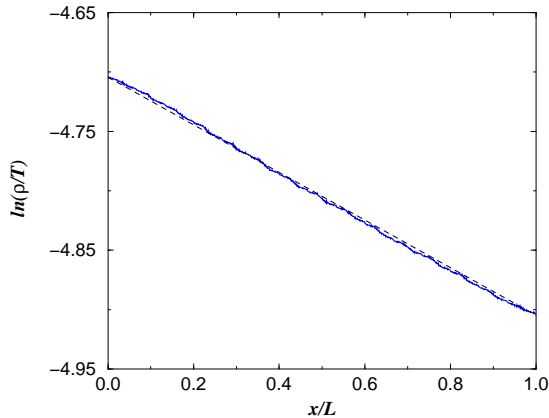}
\caption{\label{fig:mu-profile}The quantity $\ln (\rho/T)$ (solid
line) as a function of the position $x$ computed from the density and
temperature profiles obtained in the simulation described in
Fig.~\ref{fig:T-profile}. The dashed line corresponds to a linear
profile of the chemical potential $\mu/T$ joining its nominal values
in the baths as obtained from (\ref{eq:mu}). The solid line has
been shifted from its obtained numerical value for a better
comparison.}
\end{center}
\end{figure}

We now turn to the verification that LTE holds for our system. The assumption
of LTE consists in the following: in every ``infinitesimal'' volume element
one can define thermodynamic variables in the usual way and these are related
to each other through the relations which hold for the system at equilibrium.
It should be emphasized from the outset that this assumption is never exact:
there always exist corrections of the order of the gradients imposed on the
system. However, by choosing sufficiently small gradients it is always
possible to reach a situation in which a definition of the temperature and
density variables as if the volume element was in equilibrium, leads to
reasonable values for the local chemical potential. Let us first discuss the
issue of defining the temperature: in thermal equilibrium it is well known
that the particle energies have a Boltzmann distribution. We may therefore
define temperature as the parameter characterizing this distribution, arguing
that temperature is ill-defined if the distribution is not Boltzmann.  To this
end we have computed the energy distribution of the gas of particles $P(E;x)$
as they cross a narrow slab centered at some position $x$. In
Fig.~\ref{fig:lte}, we show (in symbols), the results for $P(E;x)$ measured at
three different positions for the same simulation described in
Fig.~\ref{fig:T-profile}. At each position the distribution $P(E;x)$ is
consistent with the Boltzmann distribution, thus indicating that the gas of
particles behaves locally as if it was in equilibrium at some temperature
$T(x)$. If we determine the temperature $T(x)$ by a fit of $P(E;x)$ to the
Boltzmann distribution and compare these values with the mean energy per
particle profile of the system shown in Fig.~\ref{fig:T-profile}, we see in
Fig.~\ref{fig:lte-2} that $T(x)$ coincides with the mean energy of the
particles measured locally at the position $x$. Thus, in the steady state, a
local Boltzmann distribution is established.  Therefore, the identification of
the mean energy per particle with the local temperature in this model is
justified.

In Fig.~\ref{fig:mu-profile}, we now compare the dependence on $x$ of the
quantity $\ln(\rho/T)$ with a linear profile between the nominal values of
$\mu/T$ in the thermo-chemical baths for the same simulation. The agreement
between both curves (the profile of $\ln(\rho/T)$ has been shifted for
comparison), indicates that $\Delta\mu/T=\Delta\ln(\rho/T)$, supporting the
fact that the gas of particles inside the channel behaves locally as an ideal
gas.  By this we mean that the relation between the chemical potential, the
density and the temperature for an ideal gas in equilibrium holds good locally
to an excellent approximation in the NSS under study.

\begin{figure}[!t]
\begin{center}
\includegraphics[scale=0.4]{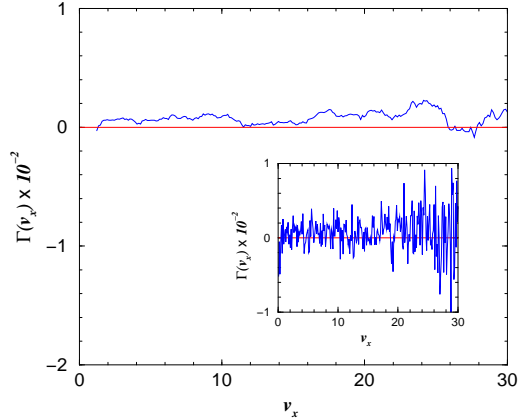}
\caption{\label{fig:lamberto}Window average of the deviation from the
Maxwell distribution given in terms of the ratio $\Gamma(v_x) =
\{P(v_x)-P(-v_x)\}/P(v_x)$ for a simulation with $\Delta(\mu/T) =
-0.06$ and $T$ constant in a channel of length $L=30$. In the Inset,
the original result for $\Gamma(v_x)$ is shown.}
\end{center}
\end{figure}

There is, however, an obvious discrepancy: since the NSS generally carries a
non-zero particle current, it is clear that the average velocity is
non-vanishing, thereby contradicting the Maxwellian distribution for the
velocities. This is shown in detail in Fig.~\ref{fig:lamberto} where the
deviation from the Maxwell distribution given in terms of ratio $\Gamma(v_x) =
\{P(v_x)-P(-v_x)\}/P(v_x)$ is plotted. The systematic positive value of
$\Gamma(v_x)$ is an indication that the average velocity is greater than zero
\cite{lamberto}. This velocity, however, is proportional to the particle
current, and hence to the gradients. Since these must be assumed small for LTE
to hold, it is a small effect, which vanishes in the relevant limit. At
this point, it is worthwhile to make the following point, when one states that
a model such as that defined in \cite{casati84} does not satisfy LTE, it means
that the deviations from LTE do not decrease as the gradients. In that case,
for example, they only decay as the imposed temperature {\em difference\/}
goes to zero, which in the thermodynamic limit is a much more stringent
condition.

\section{Normal transport and Onsager reciprocity relations}
\label {sec:transport}

Having shown that we can assign an unambiguous meaning to local
thermodynamical quantities in the non equilibrium steady states reached by the
system, we can now study the transport properties of our systems, and whether
these comply with the predictions of irreversible thermodynamics.

\begin{figure}[!b]
\begin{center}
\includegraphics[scale=0.4]{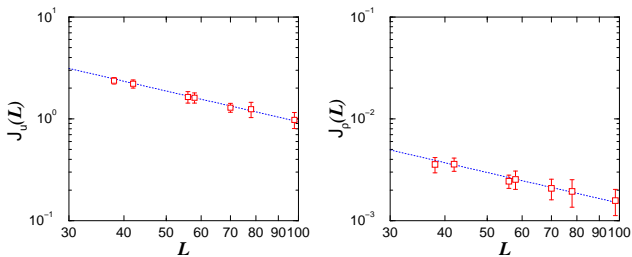}
\caption{\label{fig:normal}Size dependence of the heat and matter
currents for simulations with a fixed temperature difference, $\Delta
T=20$, and $\mu/T$ constant. The dotted lines corresponds to $1/L$
scaling.}
\end{center}
\end{figure}

From the general theory of irreversible processes, to linear order,
the heat and particle currents ${\mathrm J}_u$ and ${\mathrm J}_\rho$
can be written as follows (see e.g. \cite{mazur84}):
\begin{eqnarray}
{\mathrm J}_u&=&L_{uu}\nabla\frac{1}{T}-L_{u\rho}\nabla\frac{\mu}{T}\nonumber \ ,\\
{\mathrm J}_\rho&=&L_{\rho u}\nabla\frac{1}{T}-L_{\rho\rho}\nabla\frac{\mu}{T}
\label{eq:phenom-eqs}
\end{eqnarray}
and the Onsager reciprocity relations read in this case
\begin{equation}
L_{u\rho}=L_{\rho u} \ .
\label{eq:reciprocity}
\end{equation}
We now wish to show a central feature of our model: namely that its transport
properties are {\it normal}, meaning that the various transport coefficients
appearing in (\ref{eq:phenom-eqs}) do not depend on the length of system, and
are thus well defined in the thermodynamical limit. In Fig.~\ref{fig:normal}
we show the dependence of the currents on the length $L$ of the system, for a
typical realization, in which we keep the differences $L\nabla T$ and $L\nabla
\mu/T$ fixed as we vary the system size. The $1/L$ dependence observed
confirms that transport is normal.

In order to obtain the value of the coefficients in (\ref{eq:phenom-eqs}), it
is enough to perform two simulations: Fixing the value of $\nabla T$ and
setting $\nabla(\mu/T)=0$ yields $L_{uu}$ and $L_{\rho u}$ from the direct
measurement of the energy and particle flows, while setting $\nabla T =0$ and
fixing $\nabla(\mu/T)$ gives $L_{u\rho}$ and $L_{\rho \rho}$. We have
performed simulations with temperature and chemical potential differences up
to 20\% of the minimal nominal values at the walls, and in all cases we have
found normal transport consistent with (\ref{eq:reciprocity}).

\begin{figure}[!t]
\begin{center}
\includegraphics[scale=0.4]{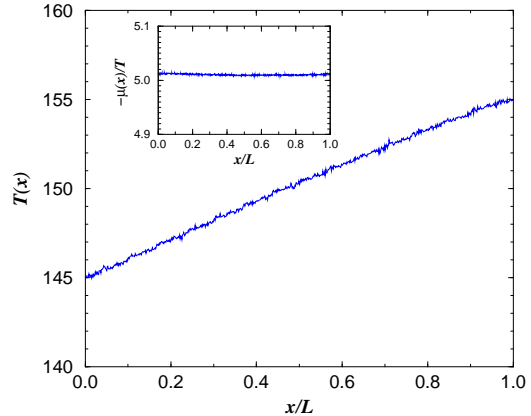}
\caption{\label{fig:mu-grad_T} Particle's temperature profile for a
simulation where the quantity $\mu/T$ is kept constant. The nominal
values for the baths' temperature are $T_0=145$ and $T_1=155$ in a
channel of length $L=30$. In the inset, the profile of $\mu/T$ as
calculated from (\ref{eq:mu}) with $\mu_0=1$ is shown to corroborate
its constant value along the channel.}
\end{center}
\end{figure}

In Fig.~\ref{fig:mu-grad_T}, we show the particle's temperature and chemical
potential profiles obtained from a simulation with constant $\mu/T$ and a
temperature difference of $\Delta T = 10$ around $T=150$ in a channel of
length $L=30$ averaged over 470 realizations. After the steady state has been
reached, we found that both a heat current and a particle current were driven
by the temperature gradient. From (\ref{eq:phenom-eqs}) we obtained the
following values for the Onsager coefficients:
\begin{eqnarray} \label {eq:L-mu}
L_{uu} & = & (0.7920 \pm 0.0050)\rho T^{5/2} \nonumber \\
L_{\rho u} & = & (0.1271 \pm 0.0017)\rho T^{3/2} \ .
\end{eqnarray}

\begin{figure}[!t]
\begin{center}
\includegraphics[scale=0.4]{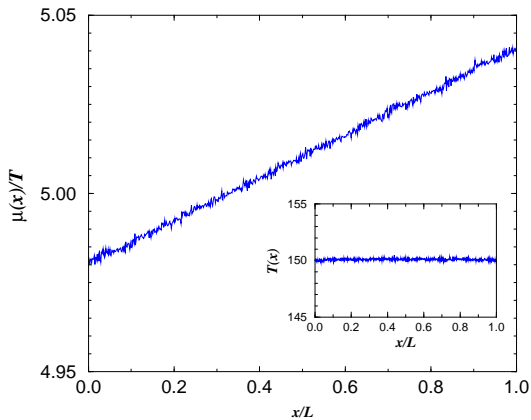}
\caption{\label{fig:T-grad_mu} Profile of the quantity $\mu/T$ as
calculated from (\ref{eq:mu}) with $\mu_0=1$ for a simulation where
the nominal values of the baths' temperature are set to the same value
$T=150$. with an end-to-end chemical potential difference of
$\Delta(\mu/T) = -0.06$ in a channel of length $L=30$. In the inset,
the particle's temperature profile is shown.}
\end{center}
\end{figure}

From the complementary simulation (see Fig.~\ref{fig:T-grad_mu}),
where temperature is kept constant at $T=150$ and a chemical potential
gradient is imposed with $\Delta(\mu/T) = -0.06$ we obtained
\begin{eqnarray}\label {eq:L-T}
L_{u\rho} & = & (0.1272 \pm 0.0048)\rho T^{3/2} \nonumber \\
L_{\rho \rho} & = & (0.1050 \pm 0.0030)\rho T^{1/2} \ .
\end{eqnarray}

\begin{figure}[!t]
\begin{center}
\includegraphics[scale=0.4]{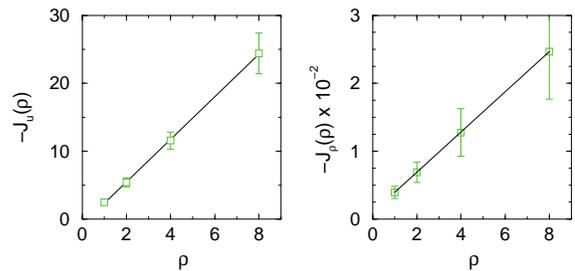}
\caption{\label{fig:density-scaling} Density dependence of the
particle ${\mathrm J}_{\rho}$ and heat ${\mathrm J}_u$ currents. Each
symbol in these plots correspond to the averaged fluxes obtained from
a simulation with constant $\mu/T$ and a temperature gradient for
which $T_0=145$ and $T_1=155$ in a channel of length $L=30$. The lines
correspond to linear fits.}
\end{center}
\end{figure}

In (\ref{eq:L-mu}) and (\ref{eq:L-T}) we have written the explicit dependence
of the Onsager coefficients on the particle's density $\rho$ and temperature
$T$. The temperature dependences arise from simple dimensional analysis given
the fact that the system is homogeneous in energy, or equivalently, that it
has no proper time scale. The same cannot be said of the density, but the
obtained linear dependence shown in Fig. \ref{fig:density-scaling}, at least
in this range of density values, is not too surprising.

The obtained values for the symmetric coefficients $L_{\rho u}$ and
$L_{u\rho}$ confirm (\ref{eq:reciprocity}) to within our numerical accuracy.

As a consistency check, we have also studied a ``canonical'' situation, in
which we suppressed absorption and emission of particles at the walls while
still allowing heat exchange. In this situation there is no flow of matter in
the steady state. The relationship between heat flow and temperature gradient
becomes ${\mathrm J}_u=\kappa \nabla T$, with $\kappa$ given by the following
expression:
\begin{equation}
\kappa=\frac{L_{uu}L_{\rho\rho}-L_{u\rho}L_{\rho u}}{T^2L_{\rho\rho}} \ .
\label{eq:canonical}
\end{equation}
This relationship was found to hold to good accuracy, thus confirming
the validity of our ``grand canonical'' simulations by which the $L$'s
were evaluated.
We remark that the coupling between the two currents in the NSS is non-trivial
in the following sense: in a ``canonical'' simulation, the simplest assumption
for the dependence of the density on the position is that the trajectory of
each particle covers the sample uniformly. Then the local temperature merely
determines the speed at which the orbit is being traversed. This would imply
that in such a situation, the particle density should scale inversely with the
average velocity, that is
\begin{equation}
\rho(x)T(x)^{1/2}=const \ .
\label{eq:inconstancy}
\end{equation}

In terms of the transport coefficients defined in
(\ref{eq:phenom-eqs}), (\ref{eq:inconstancy}) is equivalent to
$L_{\rho u} = (d+1) T L_{\rho\rho}/2$, where $d$ is the dimension of
the system. This relation corresponds to a system for which all
transport arises from uncorrelated Markovian motion of the particles,
as in the Knudsen gas \cite{reichl88}.  However,
(\ref{eq:inconstancy}) does not hold in our system as we always find
a small but systematic spatial variation in this quantity, the size
and sign of which depend on the value of $\eta$. Thus, our system
cannot be accurately described in such a simple manner.

Despite the fact  that the dynamics of the system  is not ergodic when
an external  magnetic field is  applied, we have  performed systematic
simulations for  several values of the magnetic  field. In particular,
we have measured both the  matter and the energy currents appearing in
the direction  perpendicular to the  applied thermodynamical gradients
(the so-called Righi-Leduc effect, which  is the thermal analog of the
Hall effect  \cite{mazur84}).  Even when a magnetic  field is applied,
the existence of the up-down symmetry in the system makes the dynamics
equivalent to a situation  where time-reversal symmetry holds.  Due to
this  situation  we  are  not  able  to verify  the  validity  of  the
Onsager--Casimir relations which are the generalization of the Onsager
relations when time-reversal symmetry is broken. Nevertheless, we have
verified that all cross $L$-coefficients satisfy the Onsager relations
to within  numerical accuracy.   The validity of  the Onsager--Casimir
relations deserves further investigation.

\section{Green-Kubo formalism}
\label {sec:green-kubo}
Another interesting question we are able to address with our system is whether
the Green-Kubo relations hold. These relate the phenomenological collective
transport coefficients with the equilibrium time correlation functions of
microscopic dynamical variables.

As finite size effects limit the range of validity of the Green-Kubo
relations, we begin by presenting a derivation of these relations which is
appropriate to our finite length system \cite{self-diffusion}. Starting from
the phenomenological transport equations (\ref{eq:phenom-eqs}), and using the
fact that the gas of particles is an ideal gas, we express
(\ref{eq:phenom-eqs}) in terms of the gradients of the energy density $u$ and
particle density $\rho$:
\begin{eqnarray} \label{eq:phenom-eqs-2}
{\mathrm J}_u&=&\frac{1}{\rho T}\left(L_{u\rho}-\frac{L_{uu}}{T}\right)\nabla{u} + \frac{1}{\rho}\left(\frac{L_{uu}}{T}-2L_{u\rho}\right)\nabla{\rho} \nonumber \ ,\\ 
\\
{\mathrm J}_\rho&=&\frac{1}{\rho T}\left(L_{\rho\rho}-\frac{L_{\rho u}}{T}\right)\nabla{u} + \frac{1}{\rho}\left(\frac{L_{\rho u}}{T}-2L_{\rho\rho}\right)\nabla{\rho} \nonumber \ . 
\end{eqnarray}

As the analysis is restricted to the linear regime, the factors of $\rho$ and
$T$ appearing in the coefficients of the gradients are taken as constants.
Using energy and mass conservation, we obtain from (\ref{eq:phenom-eqs-2})
\begin{equation}
\frac{\partial}{\partial t}\left(
\begin{matrix}
u \\
\rho \\
\end{matrix}
\right) = \mathbb{A} \nabla^2 \left(
\begin{matrix}
 u \\
\rho \\
\end{matrix}
\right) \ ,
\label{eq:g-k-1}
\end{equation}
where the matrix $\mathbb{A}$ is the matrix of coefficients whose
elements can be readily identified from (\ref{eq:phenom-eqs-2}).

Applying a Fourier transform in space to (\ref{eq:g-k-1}) we obtain
\begin{equation}
\left(
\begin{matrix}
\hat{u}_k(t) \\
\hat{\rho}_k(t) \\
\end{matrix}
 \right) = e^{-tk^2\mathbb{A}} \left(
\begin{matrix}
\hat{u}_k(0) \\
\hat{\rho}_k(0) \\
\end{matrix}
\right) \ , 
\label{eq:g-k-2}
\end{equation}
where $\hat{u}_k(t)$ and $\hat{\rho}_k(t)$ are the $k$ Fourier
components of $u(x,t)$ and $\rho(x,t)$ evaluated at time $t$. Now
we take the exterior product of (\ref{eq:g-k-2}) with the vector
$(\hat{u}_k^*(0) \ \hat{\rho}_k^*(0)$ and average over the ensemble
of equilibrium realizations to obtain
\begin{equation}
{\mathbb C}_k(t) = e^{-tk^2{\mathbb A}}{\mathbb U}_{k0} \ .
\label{eq:g-k-3}
\end{equation}
The matrices ${\mathbb C}_k(t)$ and ${\mathbb U}_{k0}$ consist of
correlation functions of the densities given by
\begin{equation}
{\mathbb C}_k(t) = \left(
\begin{matrix}
\left<\hat{u}_k(t)\hat{u}_k^*(0)\right> & \left<\hat{u}_k(t)\hat{\rho}_k^*(0)\right> \\ 
\left<\hat{\rho}_k(t)\hat{u}_k^*(0)\right> & \left<\hat{\rho}_k(t)\hat{\rho}_k^*(0)\right> \\
\end{matrix} 
\right)
\label{eq:g-k-4}
\end{equation}
and
\begin{equation}
{\mathbb U}_{k0} = \left(
\begin{matrix}
\left<|\hat{u}_k(0)|^2\right> & \left<\hat{u}_k(0)\hat{\rho}_k^*(0)\right> \\ 
\left<\hat{\rho}_k(0)\hat{u}_k^*(0)\right> & \left<|\hat{\rho}_k(0)|^2\right> \\
\end{matrix} 
 \right) \ .
\label{eq:g-k-5}
\end{equation}

The above correlation functions were derived explicitly as forward time
correlation functions; however, as they pertain to the stationary state of the
system, we can extend them backwards in time as even functions. Having done
this, the Fourier transform in time of (\ref{eq:g-k-3}) yields
\begin{equation}
\tilde{\mathbb C}_k(\omega)
= \frac{2k^2}{{\mathbb A}^2k^4 + \omega^2}{\mathbb A}
{\mathbb U}_{k0} \ .
\label{eq:g-k-6}
\end{equation}
If we use again the continuity equations we can express the matrix
$\tilde{\mathbb C}_k(\omega)$ in terms of time correlations of the
energy and mass fluxes as
\begin{equation}
\tilde{\mathbb C}_k(\omega) = \frac{k^2}{\omega^2} \tilde{\mathbb J}_k(\omega)
\ , 
\label{eq:g-k-7}
\end{equation}
where the matrix $\tilde{\mathbb J}_k$ is given by
\begin{equation}
\tilde{\mathbb J}_k(\omega) = \left(
\begin{matrix}
\tilde{\mathrm C}_{k{\mathrm J}_u{\mathrm J}_u}(\omega) & \tilde{\mathrm C}_{k{\mathrm J}_u{\mathrm J}_\rho}(\omega)\\
\tilde{\mathrm C}_{k{\mathrm J}_\rho {\mathrm J}_u}(\omega) & \tilde{\mathrm C}_{k{\mathrm J}_\rho {\mathrm J}_\rho}(\omega) \\
\end{matrix} 
\right) \ .
\label{eq:g-k-8}
\end{equation}
and $\tilde{\mathrm C}_{k{\mathrm J}_a{\mathrm J}_b}(\omega) = \left<\tilde{\mathrm
 J}_{ka}(\omega)\tilde{\mathrm J}_{kb}(-\omega)\right>$ are flux
correlation functions in the $k$-$\omega$ space. Inserting (\ref{eq:g-k-7})
into (\ref{eq:g-k-6}) we obtain
\begin{equation}
\tilde{\mathbb J}_k(\omega)
= \frac{2\omega^2}{{\mathbb A}^2k^4 + \omega^2}{\mathbb A}
{\mathbb U}_{k0} \ .
\label{eq:g-k-9}
\end{equation}
In the limit $k^2\to 0$, the relation in (\ref{eq:g-k-9}) becomes independent
of $k$ and $\omega$:
\begin{equation}
{\mathbb A} = \frac{1}{2} {\mathbb J}_k(\omega){\mathbb U}_{k0}^{-1} \ .
\label{eq:g-k-10}
\end{equation}
However, for finite $k$, the limit attained by taking $\omega\to 0$ yields
$\tilde{\mathbb J}_k(\omega)=0$. In (\ref{eq:g-k-10}) the Onsager coefficients
are expressed as a function of correlation functions of energy and particle
fluxes and the static correlation functions of the densities. Furthermore,
from the equilibrium statistics of the system for the energy and particle
distributions of the gas of particles, the matrix ${\mathbb U}_{k0}$ can be
explicitly written as
\begin{equation}
{\mathbb U}_{k0} = \left(
\begin{matrix}
2NT^2 & NT \\
 NT & N \\
\end{matrix}
 \right) \ .
\label{eq:g-k-11}
\end{equation}
where $N$ and $T$ are the number of particles and temperature of the
gas respectively.

Finally, inserting (\ref{eq:g-k-11}) into (\ref{eq:g-k-10}) we
obtain a compact expression for the Green-Kubo relations for the
Onsager coefficients
\begin{equation}
L_{ab} = \frac{1}{2L} \tilde{\mathrm C}_{k{\mathrm J}_a{\mathrm J}_b}(\omega) \,
\label{eq:g-k-12}
\end{equation}
for $a,b = \{u \ \rho\}$. Here we have used that $\rho L = N$.

In order to obtain a value for the Onsager coefficients from the Green-Kubo
formulas of (\ref{eq:g-k-12}) we measured the equilibrium correlation
functions of spatial Fourier components of the energy and particle currents.
To this effect, we performed the measurements in a channel composed of $L =
120$ unit cells. These measurements were done in microcanonical simulation
with periodic boundary conditions, as it is clear that transport coefficients
vanish in the static limit if the boundary conditions are reflecting. Also,
the above relations were obtained strictly as the limit of $k\to 0$, we
therefore worked at small finite values for the wave number $k$ at frequency
$\omega$, which corresponds to the situation in which (\ref{eq:g-k-9}) is
applicable. Note that the usual way of evaluating the diffusion constant
by integrating over the current correlations in the periodic case, which
corresponds to having $\omega$ small and $k$ strictly equal to zero, does not
carry over in a straightforward way when the particles are interacting, as is
the case in our system.

\begin{figure}[!t]
\begin{center}
\includegraphics[scale=0.4]{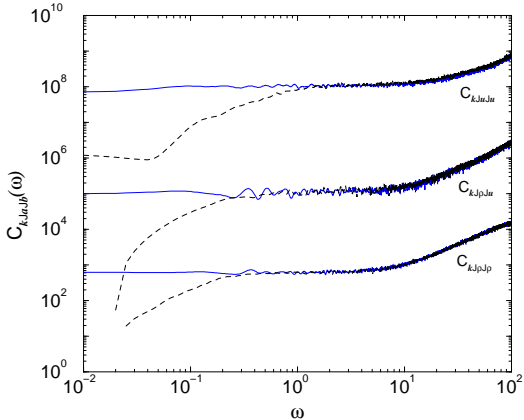}
\caption{\label{fig:green-kubo} Dependence of the different
flux-correlation functions on $\omega$. The solid lines correspond to
$\tilde{\mathrm C}_{k{\mathrm J}_a{\mathrm J}_b}(\omega)$ for
$k=1$. The dashed lines are for $k=10$. The dependence on omega of
$\tilde{\mathrm C}_{k{\mathrm J}_u{\mathrm J}_\rho}(\omega)$ is the
same that for $\tilde{\mathrm C}_{k{\mathrm J}_\rho {\mathrm
J}_u}(\omega)$ within the numerical accuracy and therefore, not
shown.}
\end{center}
\end{figure}

The periodic microcanonical simulation was performed at an energy
corresponding to a temperature $T=150$ with $N=658$ particles. The
particles and discs were set up in an equilibrium state, and we
allowed the system to further equilibrate by itself before we started
the measurements. The energy and particle flux correlation functions
were obtained as follows: We measured at each of the $n=1 \ldots L$
cells the energy ${\mathrm J}_{nu}(t)$ and particle ${\mathrm J}
_{n\rho}(t)$ currents as a function of time. They were measured at
discrete values of time with $\Delta t = 0.005$ (which corresponds to
approximately a third of the typical time required to traverse the
shortest collision path in the system). From these, we obtained the
$k$ Fourier component of the fluxes as
\begin{subequations}
\label{eq:g-k-13}
\begin{eqnarray}
{\mathrm J}_{k\rho}(t) & = & \sum_{n=1}^L e^{i2\pi kn/L}{\mathrm J}_{n\rho}(t) \\
{\mathrm J}_{ku}(t) & = & \sum_{n=1}^L e^{i2\pi kn/L}{\mathrm J}_{nu}(t) \ .
\end{eqnarray}
\end{subequations}

From a time series of $4\times 10^7$ data, approximately equivalent to
$10^6$ times the mean free flight time, we obtained the time averaged
Fourier component of the energy and particle flux time-correlation
functions
\begin{equation}
{\mathrm C}_{k{\mathrm J}_a{\mathrm J}_b}(t) = \left<{\mathrm J}_{ka}(t) {\mathrm J}_{kb}^*(0) \right>_t \ .
\end{equation}
Finally, we performed the time Fourier transform of these quantities.
In Fig.~\ref{fig:green-kubo} we show the plots of the correlation
functions ${\mathrm C}_{k{\mathrm J}_a{\mathrm J}_b}(w)$, as functions
of $\omega$ for values of $k=1$ and $10$. These plots show that at low
frequencies the correlations functions tend to zero, in agreement with
(\ref{eq:g-k-9}). However, there is a clear plateau in the plots
corresponding to the regime in which the Green-Kubo relations hold.
At higher frequencies, the curves again deviate from the plateau. This
is due to the fact that these high frequencies correspond to times
that are shorter than typical collision times, for which the diffusive
transport assumptions used to derive (\ref{eq:g-k-9}) no longer
hold. Taking the plateau values for the correlation functions, which
are essentially independent of $k$ and $\omega$, we can compute the
Green-Kubo predictions for the Onsager coefficients using
(\ref{eq:g-k-12}). In Table~\ref{tab:g-k}, we summarize the values
obtained for the different Onsager coefficients and compare them with
those obtained previously at this temperature and density.

\begin{table}
\begin{ruledtabular}
\begin{center}
\begin{tabular}{lll}
 & Green-Kubo & Gradients \\
\hline
$L_{\rho\rho}$ & $0.1050 \pm 0.003$ & $0.1030 \pm 0.002$ \\
$L_{\rho u}$ & $0.1276 \pm 0.0016$ & $0.1271 \pm 0.0017$ \\
$L_{u \rho}$ & $0.1276 \pm 0.0016$ & $0.1272 \pm 0.0048$ \\
$L_{uu}$ & $0.7920 \pm 0.012$ & $0.7710 \pm 0.005$ \\
\end{tabular}
\end{center}
\end{ruledtabular}
\caption{\label{tab:g-k} Onsager coefficients calculated from the
Green-Kubo formulas (\ref{eq:g-k-12}) compared with those values
obtained previously from direct measurements.}
\end{table}

An interesting issue involving the linear-response of the system
is the macroscopic equivalence between transport processes driven by
thermodynamical forces (due to gradients in thermodynamical
quantities) and those driven by real forces (like externally applied
fields).

It has been argued by van Kampen \cite{vanKampen71} that the linear
response theory derivation of Green-Kubo formula for the electrical
conductivity is not correct as the response of the microscopic
trajectories to an applied electric field can not be taken as linear
(see also \cite{visscher74, dorfman99}). It has been argued by Visscher
\cite{visscher74}, however, that a difference may exist between the case 
of mechanical forces, for which van Kampen's objection might hold, 
and thermodynamical forces such as temperature and chemical potential 
gradients, for which linear response should provide correct answers. 
In order to test this point, we performed simulations for our system in 
an electric field (all particles having the same charge and no interaction)
and compared this with the result of the corresponding gradient in 
chemical potential. 

At  the level  of  the Green-Kubo  relations  the equivalence  follows
directly  from the linear  response results  for a  chemical potential
gradient  or  an applied  electric  field:  Both  are related  to  the
autocorrelation function  of the  particle current when  all particles
have the same charge, which is the case in our system.

In  order to  test these  ideas  we have  performed two  complementary
simulations:  In a  first simulation  we impose  a  chemical potential
gradient  at  constant  temperature.   In  a  second  simulation  both
temperature  and  chemical  potential  are constant  and  an  external
uniform electric field is applied to the system in which the particles
now carry a unit electric charge $q$. To compare the macroscopic flows
obtained in these simulations, the magnitude of the electric field $E$
is fixed to
\begin{equation} \label{eq:E-field}
\Delta \mu = qEL \ ,
\end{equation}
so that, the work done by the electric field in taking a particle from
one  side of the  channel to  the other  is the  same as  the chemical
potential difference.  In (\ref{eq:E-field}), $L$  is the size  of the
system.

\begin{figure}[!t]
\begin{center}
\includegraphics[scale=0.4]{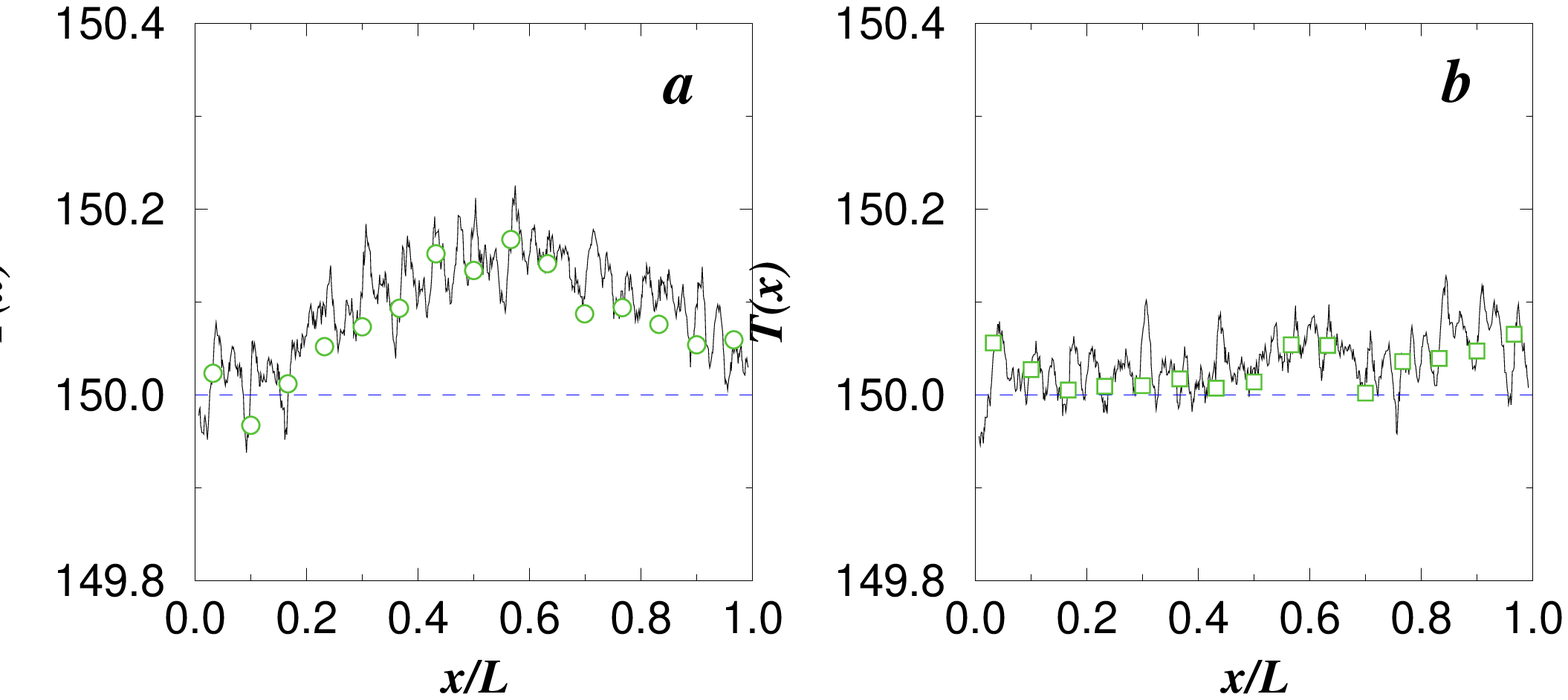}
\caption{\label{fig:E-field-T}  Temperature profiles  obtained  from a
simulation  at constant $T=150$  and mean  number of  particles $N\sim
41.2$ in a channel of length $L=30$ obtained a) from a simulation with
constant $\mu$ and a applied electric field of magnitude $E=0.5$.  and
b) without  electric field and an imposed  chemical potential gradient
of $\nabla  (\mu/T) =  0.1$.  The circles  correspond to  the averaged
kinetic energy of the discs  for the respective simulation. The dashed
line is the expected value for the profile.}
\end{center}
\end{figure}

In   Fig.~\ref{fig:E-field-T},  the   temperature   profile  of   both
simulations  is   compared.   When  the  electric   field  is  applied
(Fig.~\ref{fig:E-field-T}-a),  the  temperature  in  the bulk  of  the
system increases due  to the internal dissipation of  the work done by
the  field.   This  is  the  well known  \emph{Joule  heating}  effect
\cite{mazur84}.  In this case, we  have checked that the dependence of
the increment  of temperature in the  bulk $\Delta T$  is quadratic in
the  field and  observed  that it  holds  even far  beyond the  linear
regime.   In  the  case  of  an imposed  chemical  potential  gradient
(Fig.~\ref{fig:E-field-T}-b),  the temperature  profile appears  to be
flat,  indicating that  in this  case  no effect  equivalent to  Joule
heating is present.

\begin{figure}[!t]
\begin{center}
\includegraphics[scale=0.4]{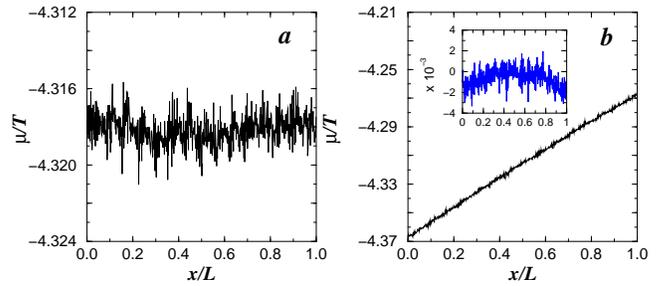}
\caption{\label{fig:E-field-rho}  Profiles  for  the quantity  $\mu/T$
minus  the expected  profile $\mu_0/T$  obtained from  the simulations
described  in Fig.~\ref{fig:E-field-T}  with (a)  an  applied electric
field and (b) with an imposed chemical potential gradient.}
\end{center}
\end{figure}

We have  also measured the  particle density profile and  obtained the
profile  for   the  quantity  $\mu/T$  computed   using  the  relation
(\ref{eq:mu})    in    both     cases.     These    are    shown    in
Fig.~\ref{fig:E-field-rho}-a for the case of an applied electric field
and in Fig.~\ref{fig:E-field-rho}-b  for an imposed chemical potential
gradient. In case (a) the chemical  potential is flat since we are not
taking into account the electric contribution. In case (b) the profile
is  linear and shows  quadratic deviations  from linearity  similar to
those observed for the temperature  profile in case (a). This is shown
in the  inset of Fig.~\ref{fig:E-field-rho}-b where  we subtracted the
theoretical linear $\mu/T$  profile.

Finally, for the  heat and matter flows we have  obtained for the case
of an applied electric field
\begin{equation} \label{eq:vanKampen-1}
\begin{array}{rcl}
{\mathrm J}_\rho &=& -0.00829 \pm 0.00006 \\
{\mathrm J}_u &=& -1.59 \pm 0.02 \ ,
\end{array}
\end{equation}
and for the case of an imposed chemical potential gradient
\begin{equation} \label{eq:vanKampen-2}
\begin{array}{rcl}
{\mathrm J}_\rho &=& -0.00829 \pm 0.00008 \\
{\mathrm J}_u &=& -1.59 \pm 0.04 \ .
\end{array}
\end{equation}
Thus, the fluxes obtained  in both simulations corroborate our initial
expectations within the numerical accuracy.

\section{Rates of Entropy Production and Phase Space Volume Contraction}
\label{sec:entropy}

It is generally argued that stationary states are characterized by minimal
rate of entropy production. In general, this rate has been at the focus of
considerable interest. In particular, in recent work on thermostated systems,
the average rate of phase space volume contraction has been identified with
the rate of entropy production \cite{chernov93, holian87, moran87,
  chernov93-2, ruelle96, dorfman99}.  While this work is certainly of
considerable interest, it is vital to be able to understand how such claims
might generalize to purely Hamiltonian systems, for which phase space volume
is rigorously conserved, at least in the fine-grained sense, due to
Liouville's theorem. Our system is well-suited to this purpose, since its
dynamics is quite transparent and its equilibrium behaviour is trivial.

In our system, it is clear that phase space volume can only be created or
annihilated at the boundaries, due to the conservation of phase space
throughout the internal dynamics (note that the collision rules
(\ref{eq:collision}) are volume preserving). To estimate the variations in
phase space volume induced by the incoming and outgoing particles, we proceed
as follows: let us denote by $T_1$ and $T_2$ the two temperatures and by
$\mu_1$ and $\mu_2$ the two chemical potentials at either end of the system
(once again, due to LTE there is a local ``temperature'' and ``chemical
potential'' in the stationary state of the system). A particle entering at one
end of the system therefore contributes to the change in phase space volume
$\Omega$ by changing both the the local energy per unit volume and the local
particle density. Since, as follows from the assumption of LTE, the local
thermodynamical variables are connected to each other in the usual way, the
increase in phase space volume is given by

\begin{equation}
\Delta\ln\Omega =\sum\limits_{i=1}^2\left(\frac{1}{T_i}\left(\Delta U\right)_i
-\frac{\mu_i}{T_i}\left(\Delta N\right)_i\right),
\label{eq:s-increase}
\end{equation}
where $i$ can take the values $1$ and $2$ for either end. Due to stationarity,
however, it is clear that

\begin{equation}
\begin{array}{lclcl}
\left(\Delta N\right)_1 & = & -\left(\Delta N\right)_2 & = & {\mathrm j}_\rho \Delta t \\
\left(\Delta U\right)_1 & = & -\left(\Delta U\right)_2 & = & {\mathrm j}_u\Delta t \ .
\end{array}
\label{eq:flow}
\end{equation}

Putting (\ref{eq:flow}) into (\ref{eq:s-increase}), one obtains 
the expression for the rate of production of the phase space volume:
\begin{equation}
\frac{d\ln\Omega}{dt}=L\left[
-{\mathrm j}_u\nabla\frac{1}{T}+{\mathrm j}_\rho\nabla\frac{\mu}{T}
\right].
\label{eq:s-prod}
\end{equation}
Using (\ref{eq:phenom-eqs}), and the positive definiteness of the matrix of
transport coefficients $L_{\alpha\beta}$, one finds that the r.h.s of
(\ref{eq:s-prod}) is always negative: on the whole, phase space volume flows
out of the system into the baths. This can be interpreted in two complementary
ways. At first, one might argue that if a system has ever contracting phase
space volume, it must eventually end up on an invariant set of measure zero.
This is in fact quite reasonable and in good agreement with various findings
on the dynamics of reversible thermostated systems. Indeed, various people
\cite{dorfman99, gallavotti95, ruelle99, evans00} have found that the
invariant measure for such systems is singular with respect to Liouville
measure. If we were to generalize such an assumption to our system, it would
mean that the invariant measure of the stationary state it eventually reaches
is also singular with respect to Liouville measure, which would in turn
signify that the total phase space volume of the steady state is zero. In
principle, this seems amenable to a numerical test, but this is not really the
case: due to the rather high dimensionality of our system, the fractality of
the support of the steady state measure is presumably extremely difficult to
observe, if it is at all possible. Therefore, even if we could connect entropy
production, say, with some kind of escape rate (as proposed by Gaspard and
coworkers \cite{gaspard98}), it is not at all clear how to measure such a
quantity in even so simple a model as ours. Whether truly low-dimensional
systems exist, for which such quantities are readily accessible and which also
exhibit LTE and normal transport, is an interesting open problem.

Another interpretation of (\ref{eq:s-prod}) is the following: in the
stationary state, because we are in fact in LTE, that is, approximately in a
state of thermal equilibrium in every volume element, we may think that at
least the {\em coarse grained} phase space volume, measured by the
thermodynamical entropy, will remain constant in steady state. But if, as
claimed in (\ref{eq:s-prod}), an amount of phase space volume leaves the
system each unit of time, then an equivalent amount of coarse-grained phase
space volume must be produced inside the system. That this last statement does
not contradict Liouville's theorem is, of course, well-known from traditional
examples in statistical mechanics. Such an interpretation might then justify
the interpretation of the r.h.s of (\ref{eq:s-prod}) as minus the rate of
entropy production, in conformity with the results stated for reversible
thermostated systems.

To make such an identification plausible, we need a definition of the entropy
out of equilibrium. It is well-known that in the general case such a
definition poses formidable problems. However, since our system is an ideal
gas but little perturbed from equilibrium, it is possible to use Boltzmann's
expression for an entropy density per unit volume in terms of the one-particle
distribution function $f(x, v;t)$, defined by
\begin{equation}
s(x)=-\int dv\,f(x,v;t)\ln f(x,v;t)
\label{eq:s-boltz}
\end{equation}
From this one may give an elementary expression for the rate of entropy
production through the following considerations. Due to LTE, the stationary
distribution $f(x,v)$ is given by a local Maxwellian
\begin{equation}
f(x,v)=\rho(x)\beta(x)^{1/2}\exp\left[-\beta(x)v^2/2\right] + \delta f(x,v)
\label{eq:local}
\end{equation}
where $\delta f(x,v)$ is a correction term of the order of the imposed
gradients, which accounts for the currents in the system. As shown in
(\ref{fig:lamberto}) such a correction has been observed in our system. It
then seems reasonable to divide the time variation of $f(x,v)$ into a
convective part given by $-vf_x$ plus a collisional part, of which we need say
nothing. In this case, we may define the entropy current density as follows

\begin{eqnarray}
{\mathrm j}_s & = & -\int dv\,vf(x,v)\ln f(x,v) \\
 & = & -\int dv\,v\delta f(x,v)[\ln f(x,v)+1] \ . \\
\label{eq:current}
\end{eqnarray}

Inserting (\ref{eq:local}) into (\ref{eq:current}) yields

\begin{equation}
{\mathrm j}_s=-\frac{\mu}{T}{\mathrm j}_\rho+\frac{1}{T}{\mathrm j}_u \ .
\label{eq:current1}
\end{equation}

In the stationary state, the local rate of entropy production
is given by the divergence of the entropy current. If one takes
it as given by (\ref{eq:current1}), we obtain again the rate of entropy 
production given by the standard expression, this time with the correct
positive sign. 


\section{Conclusion}
\label {sec:conclusions}

We have introduced a simple model system characterized by reversible
Hamiltonian microscopic dynamics, the equilibrium properties of which are
those of an ideal gas when viewed as a thermodynamic system. We performed
extensive numerical studies of the system both in equilibrium and in
non-equilibrium steady state. We find that when driven out of equilibrium, the
properties of the system are consistent with the hypothesis of local thermal
equilibrium, that is, in the stationary state, one finds a local Boltzmann
distribution for the energies, leading therefore to an unambiguous definition
of the local temperature, chemical potential, etc. In this situation its
transport properties are found to be entirely similar to those of realistic
interacting many particle systems: it exhibits coupled mass and heat transport
and the two cross transport coefficients satisfy Onsager's relations. When
time reversal invariance is broken by means of an applied constant magnetic
field, the appropriate generalizations of the Onsager relations hold. This is
the case even though this system turns out not to be ergodic. Furthermore, we
find extremely satisfactory agreement between the values of the transport
coefficients as deduced from the equilibrium dynamics via the Green--Kubo
formulas and those obtained via direct simulation. A final result of interest
is that we were able to show equivalence between an applied electric field and
a chemical potential gradient. This result, while a straightforward
consequence of linear response, does not appear at all obvious from the
microscopic point of view. It may therefore be considered as a further
confirmation of the validity of the linear response formalism

The fact that all these features are present in such a simple model,
suggests that it may serve as an ideal framework to gain insight of how
macroscopic transport phenomena arise in real systems. The model is also well
suited to test theories for the description of systems in out of equilibrium
states. Among these, a study of the Cohen--Gallavotti theorem might be very
interesting. Also, issues linked to those involving entropy production
discussed in the section \ref{sec:entropy}, in particular the question of
characterizing the invariant measure of the stationary state appears very
challenging. 

Finally, extensions to more complicated systems can also be considered. By
varying the value of the moment of inertia of the rotors one can study
transport in heterogeneous structures, such as junctions, layered systems,
etc. Other extensions could include mixtures of scattered particles with
different masses. For all these extensions, most if not all of the equilibrium
properties are straight forward, and theories for the phenomenological
behaviour of such systems can be carefully tested.

\begin{acknowledgments}
We acknowledge enlightening discussions with E. G. D. Cohen, J. Lebowitz,
T. Prosen and L. Rondoni; as well as financial support from UNAM-DGAPA project
IN112200 and CONACYT project 32173-E. We also thank the Centro Internacional
de Ciencias AC for hospitality during a part of this work. 
\end{acknowledgments}

\appendix
\section{Various Possible Thermo-chemical Baths}
\label{app-A}
In this appendix we present a short discussion of Markov processes that can
be used to generate a canonical or grand canonical ensemble. A fairly general
Markov process combined with a Hamiltonian dynamics satisfies the master
equation
\begin{eqnarray}
\partial_t\rho_N(x)&=&\{H_N,\rho_N\}+\sum_{N^\prime}\int W(N^\prime N;
x^\prime x)\rho_{N^\prime}(x^\prime)dx^\prime-\nonumber\\
&&-\rho_N(x)\sum_{N\neq N^\prime}\int W(N N^\prime;
x x^\prime)dx'.
\label{eq:a.1}
\end{eqnarray}
Here $N$ is the number of particles, $x$ is an abbreviated notation for the
vector $(p,q)$ and $\rho_N(x)$ is the probability density of being at the
phase point $x$ having altogether $N$ particles. The $W(NN^\prime;xx^\prime)$
are the transition rates of going from a phase point $x$ with $N$ particles to
a phase point $x^\prime$ with $N^\prime$ particles. If we are looking for
rates such that they generate, say, the grand canonical ensemble with a given
temperature and chemical potential, it is sufficient that they satisfy
the detailed balance condition
\begin{equation}
\frac{W(NN^\prime;xx^\prime)}{W(N^\prime N;x^\prime x)}
=\exp[\beta\mu(N-N^\prime)-\beta(H_{N^\prime}(x^\prime)-
H_N(x))].
\label{eq:a.2}
\end{equation}
Let us now assume that the stochastic part of the dynamics is 
limited to the case in which $x$ belongs to a small subset $\Gamma$ 
of the full phase space, otherwise the dynamics is purely Hamiltonian. 
This corresponds to the case of stochastic walls treated in this 
paper. Assume further that, as in our model, the Hamiltonian in 
$\Gamma$ is one-particle only, say a pure kinetic energy term. Assume 
finally that the only changes we consider will be the introduction and 
destruction of a single particle. Then the following 
rates are a solution of (\ref{eq:a.2}):
\begin{equation}
W_{\Gamma}(N, N+1)=\lambda e^{-\beta (\frac{v^2}{2} +\mu)}\qquad
W_{\Gamma}(N+1, N)=\lambda 
\label{eq:a.3}
\end{equation}
If we want this stochastic process to act as a localized bath in the
system, we take the region $\Gamma$ in phase space defined by the
condition that at least one particle is in the region $V$ of the
(one-particle) configuration space. The rates in (\ref{eq:a.3}) then
mean that any particle entering $V$ is annihilated at a rate
$\lambda$. Further, particles of speed $v$ are being created inside
$V$ at a rate $\lambda e^{-\beta (\frac{v^2}{2} +\mu)}$. Since such
particles may be immediately reabsorbed, however, the problem of the
particle flux emitted by $V$ is not quite straightforward. If $V$ is
narrow (of thickness $l$) in one dimension around a given hypersurface
$S$, however, and $\lambda$ is ultimately made to go to infinity as
$l\to0$, the problem can be solved as follows. For simplicity, assume
that all momenta always point in the direcion of one of the sides of
$V$. The probability that the particle will come out at all,
conditioned on its having been created at a distance $r$ of the side
from which it must leave $V$ is given by
\begin{equation}
p(r)=\exp\left[-\lambda r/v_n\right],
\label{eq:a.4}
\end{equation}
where $v_n$ is the velocity normal to the surface $S$. The flux of particles
coming out of $V$ is then given by
\begin{equation}
\lambda e^{-\beta v^2/2}\int_0^l dr\,p(r)=
v_n e^{-\beta v^2/2+\beta\mu}\left(1-e^{-\alpha/v_n}\right),
\label{eq:a.5}
\end{equation}
where $\alpha$ is the limiting value of $\lambda l$. In the limit in which
$\alpha\to\infty$, this model reduces to the one described in the text, which
was the one primarily used in the simulations presented in this paper. The
above model can be generalized in different ways: in particular, we may allow
for thermalization without change in the number of particles. A possible
candidate for such a reaction rate is given by
\begin{equation}
W(NN;xx^\prime)=\lambda'\exp\left[
-\beta(v^\prime{}^2-v^2)/2
\right]\delta(q-q^\prime),
\label{eq:a.6}
\end{equation}
where again the process acts only in the region $\Gamma$ described above.
This corresponds to a process in which the particle velocities are thermalized
while leaving their positions fixed, which is, as is readily seen, the
algorithm for thermalization through collisions introduced in the text. In
particular, the case $\alpha=0$ and a finite rate of the type given is
(\ref{eq:a.6}) leads to the algorithm we used for generating the canonical
ensemble.

Various simulations were made with finite values of $\alpha$ in order to test
the method. In equilibrium it was always found that the correct nominal
temperatures and chemical potentials were attained after sufficient time. In
non-equilibrium situations, however, for small values of $\alpha$, the
thermodynamic parameters of the system can differ from the values imposed by
the baths. In particular, the energy density showed jumps localized in the
vicinity of the walls. These energy gaps have been frequently observed in
simulations of other transport models
\cite{prosen92,lepri97,lepri98,hatano99,aoki00} and were studied in
\cite{aoki01}. In our model, they eventually disappear as we increase the
value of $\alpha$, meaning that the absorption of the wall increases and the
particles (which thermalize the system with the bath) are exchanged more
easily.



\begin{thebibliography}{}
\bibitem{homogeneity} Due to the homogeneity in energy of this system
there is not no proper energy scale or equivalently, no proper time
scale. Therefore, all energies and temperatures reported in this work
are given in arbitrary units.
\bibitem{lamberto}  Actually,   the  obtained  average   velocity  was
$<\!\!v_x\!\!>  \quad =  0.004$, which  is of  the same  order  of the
particle current  ${\mathrm J}_\rho =  0.0027$. We have  also measured
the deviations from the Maxwell distribution for huge gradients in the
chemical  potential, far  from  the linear  regime.  There, while  the
deviations are much  clearer, not just in the  average velocity but in
the width  of the distribution too,  their relative value  is still of
$1\%$.
\end{thebibliography}
\end{document}